\documentclass[lettersize,journal]{IEEEtran}
\usepackage[flushleft]{threeparttable}
\usepackage{algorithmic}
\usepackage{cite}
\usepackage{array}
\usepackage[caption=false,font=normalsize,labelfont=sf,textfont=sf]{subfig}
\usepackage{etoolbox}
\usepackage{algorithm}
\usepackage[]{footmisc}
\usepackage{textcomp}
\usepackage{enumitem}
\usepackage{stfloats}
\usepackage{url}
\usepackage{verbatim}
\usepackage{graphicx}
\usepackage{multirow}
\usepackage{mathtools}
\usepackage{float} 

\usepackage[flushleft]{threeparttable}
\usepackage{multicol}
\usepackage{array,colortbl,xcolor}
\usepackage{color,soul}
\usepackage{xcolor}
\usepackage{tikz}

\usepackage{mathtools}
\usepackage{amsmath,amsfonts}
\usepackage{amssymb,mathtools}
\usepackage{dashbox}%
\usepackage{algorithm}
\begin{document}


\title{SCALE-X: A Systematic Complexity-Aware Low-Precision Approach for Digital Predistortion}

\author{Anees~Rehman,
        Mohd~Tasleem~Khan,~\IEEEmembership{Senior~Member,~IEEE,}        ~George~Goussetis,~\IEEEmembership{Fellow,~IEEE,}\\ ~Yuan~Ding,~\IEEEmembership{Member,~IEEE,}  Jaiyu Hou, and~Jo\~{a}o~F. C.~Mota,~\IEEEmembership{Member,~IEEE}
\thanks{This work was supported by the UK Engineering and Physical Sciences Research Council (EPSRC) under Grant EP/Y037197/1.}

}


\maketitle


\begin{abstract}
This letter presents the SCALE-X framework for digital predistortion, a systematic, complexity-aware, model-based approach. The proposed scheme employs a novel complexity-aware coefficient pruning method, which, when combined with optimized fixed-point modeling, enables accurate capture of quantization effects and supports better-informed decisions for implementation. Targeted configuration of model order, memory depth, and coefficient bitwidth dramatically reduces computational overhead without noticeable reduction in accuracy. Field-programmable gate array implementation demonstrates practical trade-offs with improved efficiency.
\end{abstract}


\section{Introduction}
\label{sec:intro}
\IEEEPARstart{P}{ower} amplifiers (PAs) are inherently nonlinear devices,
introducing spectral regrowth and signal distortion that degrade the performance
of modern wireless communication systems~\cite{b2}. Digital predistortion (DPD) is the
predominant linearization technique, suppressing out-of-band emissions and
restoring signal fidelity without sacrificing PA efficiency~\cite{zhu2013blind}.
With more stringent requirements in beyond-5G, DPD
must reduce the adjacent channel power ratio (ACPR) and
error vector magnitude (EVM) targets---3GPP 5G NR mandates $\geq 45$~dBc
and $3.5$--$8$\% respectively~\cite{haider2022predistortion, khan2026next,
huang2020novel}. However, deploying DPD on resource-constrained platforms such as field-programmable gate arrays (FPGAs) makes it challenging to maintain high-fidelity linearization within tight hardware budgets~\cite{li2021fpga, chen2026comet, alhartomi2023low}. Understanding how DPD model complexity scales with hardware budget is therefore essential.

Model-based approaches such as memory polynomial (MP)~\cite{morgan2006generalized}, dynamic deviation reduction~\cite{zhu2006dynamic}, and Volterra models~\cite{huang2020novel} achieve accurate linearization by capturing nonlinearity and memory effects~\cite{cappello2022power}. However, increasing polynomial order, or memory depth, or both increases the number of coefficients, and thus the computational cost\cite{morgan2006generalized, li2022fpga}. Among model-based techniques, MP-DPD offers a favorable trade-off between linearization performance and implementation complexity, making it widely adopted \cite{braithwaite2017digital}. Its coefficients are typically estimated using least-squares (LS) fitting~\cite{cheang2018hardware}, which sometimes can yield ill-conditioned values in over-parameterized models \cite{pham2018partial}. Unlike LS, constrained LS (CLS) bounds coefficient magnitudes to improve numerical stability~\cite{guan2012bandwidth}. However, the impact of CLS on fixed-point (FxP) implementation \cite{khan2023analyzing} of DPD remains largely unstudied  \cite{enzinger2017competitive}.

Model pruning~\cite{chen2014efficient} and term selection~\cite{becerra2019comparative} retain only the most significant basis functions, reducing the number of multiplication operations while preserving linearization performance. Alternative basis functions, such as splines~\cite{barradas2014polynomials} and decomposed vector rotations~\cite{zhu2015decomposed}, together with look-up table approaches~\cite{pham2018partial}, further cut computational overhead, while real-time model switching adapts active terms per sample to reduce complexity~\cite{b1}. On the implementation side, FPGA realizations achieve efficient DPD within tight resource and energy budgets~\cite{li2021fpga,cappello2022power}, and efficient multiplier designs reduce hardware cost, often at the expense of latency and accuracy~\cite{khan2024area, rathod2025power}. Yet a unified treatment of FxP quantization and coefficient pruning remains absent from the literature. To bridge this gap, we propose \mbox{SCALE-X}, a complexity-aware framework for FPGA-based DPD. The main contributions are: 
\begin{itemize}
    \item We propose an anti-diagonal coefficient pruning strategy that reduces MP model complexity while preserving linearization performance. 
    \item We develop a quantization-aware FxP model that jointly optimizes pruning and wordlength, enabling efficient accuracy--complexity trade-offs.
    \item We validate the proposed architecture on FPGA  achieving reduced logic utilization and improved overall efficiency.
\end{itemize}
The remainder of this letter is organised as follows: Section~\ref{SecI} presents the SCALE-X framework; Section~\ref{SecIV} benchmarks the design against the state-of-the-art; and Section~\ref{SecV} concludes.


\section{Proposed SCALE-X Framework}\label{SecI}
In model-based approaches, DPD is typically developed using the PA's inverse model with an indirect learning architecture~\cite{khan2026next}. Generally, the output \(y(n)\) is expressed as
\begin{equation}\label{eq1}
y(n)=\mathrm{sum}\big(\mathbf{X}(n)\circ\mathbf{C}\big),
\end{equation}
where `$\circ$' denotes Hadamard elementwise multiplication,  
$\mathbf{X}(n)=[\chi(n-m,k)]_{M\times K}$,
$\mathbf{C}=[c(m,k)]_{M\times K}$, $m=0,\ldots,M-1$, $k=1,\ldots,K$,
and for the MP model, 
\[
\chi(n-m,k)=x(n-m)|x(n-m)|^{k-1}.
\]
\noindent This requires $M(2K-1)$ multiplications (including the generation of nonlinearities) and $(MK-1)$ additions. Alternatively, \eqref{eq1} can be expressed as
\begin{equation}\label{eq2}
y(n) = \mathrm{sum}\!\left(
\boldsymbol{\chi}_0(n) \circ
\left(\sum\nolimits_{k=1}^{K} \left(|\boldsymbol{\chi}_0(n)|^{\circ(k-1)}\right) \circ \mathbf{c}_k\right)
\right),
\end{equation}
where $\circ(k-1)$ denotes the Hadamard power, $\boldsymbol{\chi}_0(n) = \{x(n-m)\}_{m=0}^{M-1}$ 
is the input vector of memory depth $M$, and $\mathbf{c}_k = \{c_{mk}\}_{m=0}^{M-1}$ 
is the coefficient vector for nonlinearity order $K$. Unlike \eqref{eq1}, 
\eqref{eq2} requires only $MK$ multiplications.

The representation~\eqref{eq2} reduces multiplications via Horner's 
rule~\cite{enzinger2017competitive}, yet both formulations retain $MK$ 
coefficients with identical normalized mean-square error (NMSE), ACPR, and EVM performance. This work therefore 
proposes a novel approach that selectively prunes the coefficient 
matrix $\mathbf{C}=[\mathbf{c}_1,\dots,\mathbf{c}_K]\in\mathbb{C}^{M\times K}$,
reducing computational cost while preserving linearization performance. 
From simulation, it is observed that the coefficients toward the 
bottom-right corner of $\mathbf{C}$, corresponding to larger memory indices 
$m$ and nonlinear orders $k$, contribute least to linearization. These coefficients are therefore removed progressively using an
anti-diagonal pruning strategy that ranks coefficients by $m+k$.
Specifically, with $m=0,\ldots,M-1$ and $k=1,\ldots,K$, the $r^{\text{th}}$
anti-diagonal to be pruned is defined as
\begin{equation}\nonumber
\mathcal{D}_r
=
\left\{(m,k):m+k=M+K-1-r\right\},
\
\label{eq:antidiag}
\end{equation}
where $r=0,1,\ldots,(M+K-2)$ and $\mathcal{D}_0$ contains the bottom-right coefficient
$c_{M-1,K}$. The cumulative set of pruned coefficients after stage
$r$ is $\mathcal{Z}_r=\bigcup\nolimits_{d=0}^{r}\mathcal{D}_d,$
\begin{figure}[t]
  \centering\includegraphics[width=1\linewidth]{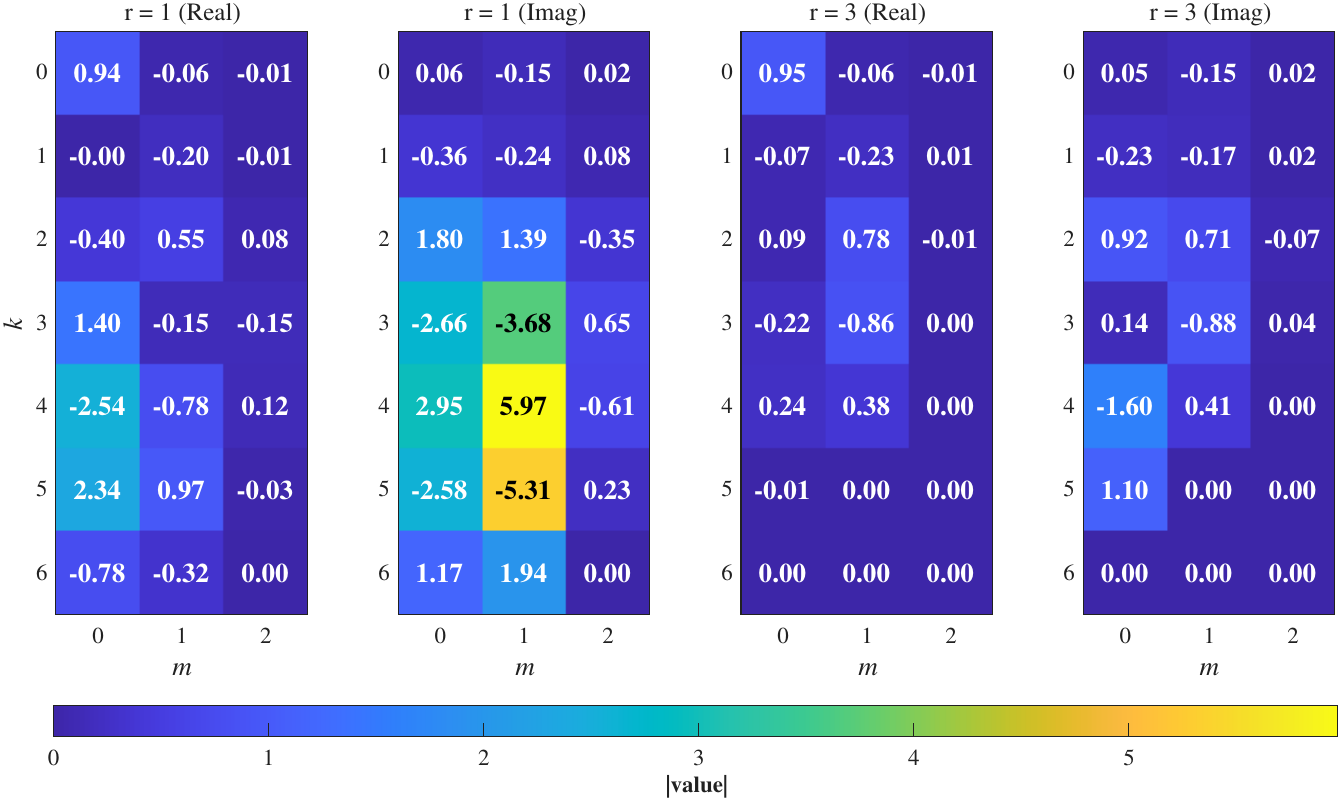}
  \caption{Illustration of coefficients' heatmap for $K=7$ and $M=3$ with $r=1$ (left) and $r=3$ (right) to demonstrate the proposed approach.
  }
  \label{fig1}
\end{figure}
and the corresponding active coefficient set is
\begin{equation}\nonumber
\mathcal{A}_r
=
\{(m,k):0\leq m\leq M-1,\;1\leq k\leq K\}
\setminus\mathcal{Z}_r.
\label{eq:activeset}
\end{equation}
Thus, the pruned coefficient matrix is
\begin{equation}\nonumber
[\mathbf{C}^{(r)}]_{mk}
=
\begin{cases}
0, & (m,k)\in\mathcal{Z}_r,\\
c_{mk}, & (m,k)\in\mathcal{A}_r.
\end{cases}
\label{eq:sparsified}
\end{equation}
At each pruning stage, the active coefficients are re-estimated using
CLS, with coefficients in $\mathcal{Z}_r$
constrained to zero, rather than simply setting the corresponding
coefficients of the original solution to zero. Specifically,
\begin{equation}
\hat{\mathbf{c}}_{\mathcal{A}_r}^{(r)}
=
\arg\min_{\mathbf{c}_{\mathcal{A}_r}}
\left\|
\mathbf{y}
-
\mathbf{\Phi}_{\mathcal{A}_r}
\mathbf{c}_{\mathcal{A}_r}
\right\|_2^2,
\label{eq:CLS}
\end{equation}
where $\mathbf{\Phi}_{\mathcal{A}_r}$ contains only the columns of the 
regression matrix associated with the active coefficient set $\mathcal{A}_r$.
The remaining coefficients in $\mathcal{Z}_r$ are fixed to zero. The resulting
$\hat{\mathbf{c}}^{(r)}$ is substituted into~\eqref{eq2} to evaluate the
linearised output, and the corresponding NMSE, ACPR, and EVM are recorded. Pruning continues until the first stage at which the linearization performance
falls outside the prescribed limits. Specifically, a pruning stage is considered
feasible when NMSE and EVM remain below their respective limits and ACPR remains
above its required minimum. If any of these performance requirements is
violated, the previous stage is selected as the maximum feasible pruning stage,
i.e., $r^*=r-1$. The resulting $\mathbf{C}^{(r^*)}$ therefore provides the
maximum pruning achieved by the proposed sequential strategy while satisfying
the specified linearization requirements. An example of the proposed approach,
with coefficient heatmaps for $K=7$ and $M=3$ at $r=1$ and $r=3$, based on the
simulation settings in Section~III, is shown in Fig.~\ref{fig1}.

\begin{figure}[t]
  \centering\includegraphics[width=1\linewidth]{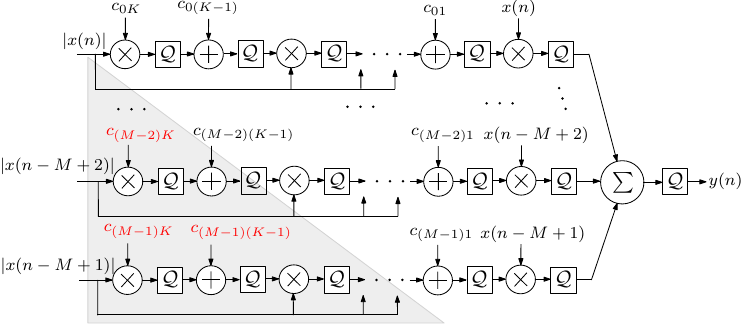}
  \caption{Proposed architecture for~\eqref{eq:horner_fixed}; pruned coefficients are in shaded area.}
  \label{fig3}
\end{figure}

For FPGA realization, inputs, coefficients, and intermediate products must be represented using FxP arithmetic. To account for finite precision, \eqref{eq2} is extended by applying an FxP quantizer to each intermediate product as follows:
\begin{equation}
y(n) = \mathcal{Q}\left[\mathrm{sum}\!\left(\mathcal{Q}\left[
\boldsymbol{\chi}_0(n) \circ
\left(\mathcal{Q}\left[\sum\nolimits_{k=1}^{K} \mathcal{Q}\!\left[ \mathbf{p}_k\right]\right]\right)
\right]\right)\right],
\label{eq:horner_fixed}
\end{equation}
where $\mathcal{Q}[\cdot]$ denotes FxP quantization in signed $\mathcal{Q}_{I.F}$ format, with 
wordlength $W = I + F$ bits comprising $I$ integer bits (including sign), and 
$F$ fractional bits. The parameters $(I, F)$ are jointly optimized alongside 
the pruning stage $r^*$, minimizing hardware complexity while maintaining 
acceptable linearization performance.


The FPGA architecture implementing \eqref{eq:horner_fixed} is shown in Fig.~\ref{fig3}. Each of
the $M$ rows corresponds to a delayed input sample $x(n-m)$, whose
magnitude $|x(n-m)|$ is raised to successive powers to form the
nonlinear basis across $K$ orders, consistent with $\mathbf{p}_k$
in~\eqref{eq2}. Within each row, basis functions are multiplied by
$c_{mk}$ and accumulated across all $K$ orders via a chain of
multiplier, quantizer and adder stages, realizing the inner sum
of~\eqref{eq:horner_fixed}; row outputs are then summed across all
$M$ taps to yield $y(n)$, with a final quantization stage at the
output. Coefficients highlighted in the shaded area correspond to the pruned
anti-diagonals $\mathcal{D}_0,\dots,\mathcal{D}_{r^*}$ of
$\mathbf{C}^{(r^*)}$, zeroed by the proposed
method. Algorithm 1 summarizes the pruning strategy and FxP optimization of the SCALE-X framework.



\begin{figure}[t]
	\centering
	\scalebox{0.85}{
		\begin{minipage}{\linewidth}
			\begin{algorithm}[H]
				\caption{Pruning and FxP Optimization}
				\begin{algorithmic}[1]
					\STATE \textbf{Input:} $x(n),y(n)$, $M,K$; \quad \textbf{Output:} $\mathbf{C}^{(r^*)}$, $(I^*,F^*)$
					\STATE \textbf{Initialize:} $\mathbf{C}^{(-1)} \gets$ CLS solution, $\mathcal{Z}_{-1} \gets \emptyset$
					\FOR{$r = 0$ to $M+K-2$}
					\STATE $\mathcal{D}_r \gets \{(m,k):m+k=M+K-1-r\}$; \ \\$\mathcal{Z}_r \gets \mathcal{Z}_{r-1}\cup\mathcal{D}_r$; \ $\mathcal{A}_r \gets \{(m,k)\}\setminus\mathcal{Z}_r$
					\STATE $\hat{\mathbf{c}}^{(r)}_{\mathcal{A}_r} \gets \arg\min\|\mathbf{y}-\mathbf{\Phi}_{\mathcal{A}_r}\mathbf{c}_{\mathcal{A}_r}\|_2^2$, \ $\mathbf{C}^{(r)} \gets [\hat{\mathbf{c}}^{(r)}_{\mathcal{A}_r},\,\mathbf{0}_{\mathcal{Z}_r}]$
					\STATE Evaluate NMSE, ACPR, EVM; \\ \textbf{if} constraint violated: $r^* \gets r-1$, \textbf{break}
					\ENDFOR
					\FOR{each $(I,F)$}
					\STATE $\mathbf{C}_{I,F} \gets \mathcal{Q}_{I,F}(\mathbf{C}^{(r^*)})$; check feasibility
					\ENDFOR
					\STATE $(I^*,F^*) \gets \arg\min_{(I,F)\ \text{feasible}} I+F$
					\STATE \textbf{return} $(\mathbf{C}^{(r^*)},I^*,F^*)$
				\end{algorithmic}
			\end{algorithm}
		\end{minipage}
	}
\end{figure}

\section{Results and Discussion}\label{SecIV}
\begin{figure}[t]
  \centering  \includegraphics[width=1\linewidth]{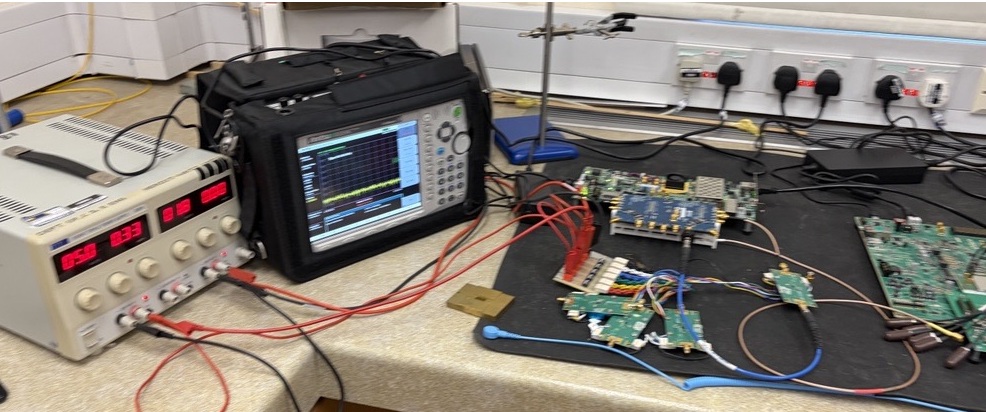}
  \caption{Experimental setup, in which a ZC706 FPGA with an FMCOMMS5 transceiver drives and captures the QPA9822 PA at 3.5\,GHz.}
  \label{fig:setup}
\end{figure}

Experiments were conducted using an in-house PA built around the Qorvo QPA9822 at 3.5\,GHz, with a measured small-signal gain of 38.24\,dB and an input P1dB of $-11.23$\,dBm. A Xilinx ZC706 FPGA platform with an FMCOMMS5 radio board generated the PA input and captured the PA output, as shown in Fig.~\ref{fig:setup}. The test signal was a 10\,MHz 5G NR waveform with a PAPR of 9.3\,dB (at 0.01\% CCDF), sampled at 40\,MSPS. The PA operated at 6.3\,dB average input back-off from P1dB, with peaks compressed by 2.8\,dB.

\subsection{Performance Evaluation}

\begin{figure*}[!t]
\centering
\includegraphics[width=7.2in]{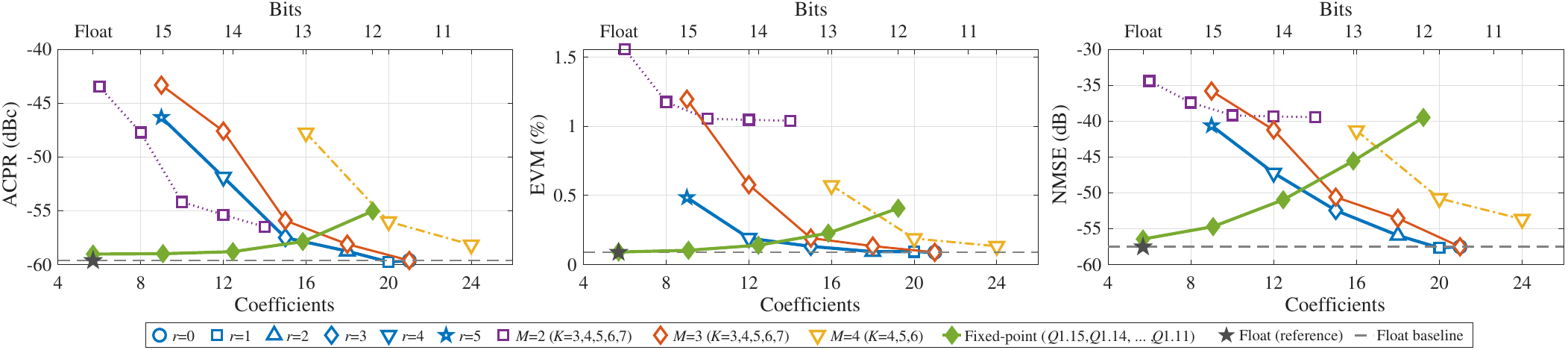}
\caption{Trade-off between model complexity and DPD linearization performance for three metrics: ACPR (left), EVM (middle), and NMSE (right); where dashed line corresponds to ideal linear output.
}
\label{fig:prev_vs_perf1}
\end{figure*}

Fig.~\ref{fig:prev_vs_perf1} shows the trade-off between model complexity and bit precision on DPD linearization performance (ACPR, EVM, NMSE). The floating-point baseline and the proposed method, under identical conditions and an equal number of active coefficients, across three memory-depth configurations, $M=2$, $M=3$, and $M=4$, each pruned progressively from $r=0$ (full model) toward higher $r$ (fewer coefficients). The FxP with bit-width shown on the top axis for direct comparison against the fixed-point curve. Among the three, $M=3$ and $K=7$ offers the best accuracy--complexity balance, sustaining near-floating-point ACPR ($-59.63$ dBc), EVM (0.088\%), and NMSE ($-57.63$ dB) down to roughly 12--14 coefficients before all three metrics degrade sharply beyond that point. $M=2$ saturates at a noticeably worse performance floor even with minimal pruning, showing that its shallow memory depth limits achievable linearization regardless of coefficient count, while $M=4$ matches or exceeds $M=3$ only at high coefficient counts and degrades sooner under pruning, reflecting redundancy among its additional memory terms.

The FxP curve, obtained by quantizing the pruned $M=3$ coefficients from Q1.15 down to Q1.11, closely tracks the floating-point pruning trend across all three metrics down to roughly 12--13 bits, indicating that reducing precision over this range costs little beyond what pruning alone already costs. Below this point, quantization noise begins to dominate and the FxP curve diverges from the floating-point baseline more sharply than coefficient count alone would predict, marking the point where precision, rather than model complexity, becomes the binding constraint on performance. This crossover indicates a practical joint operating point, roughly 12--14 coefficients at 12--13 bits, where pruning and FxP quantization can be combined with negligible compounding loss, directly motivating the SCALE-X co-design strategy of optimizing complexity and precision together rather than sequentially.

Fig.~4 compares the two complexity-reduction strategies, anti-diagonal coefficient pruning (blue) and direct $M\times K$ matrix-size reduction (orange), at matched coefficient counts across the following cases: (a) $r=5$ vs. $3\times3$, (b) $r=4$ vs. $3\times4$, (c) $r=3$ vs. $3\times5$, and (d) $r=2$ vs. $3\times6$, with the PA output without DPD (black) shown as reference. Both strategies substantially suppress adjacent-channel spectral regrowth relative to the no-DPD case, but anti-diagonal pruning consistently achieves lower out-of-band PSD than matrix-size reduction, particularly in the near-band region ($\pm$5--10 MHz), where the orange trace sits above the blue trace across all four sub-plots. This gap is largest at higher coefficient counts (a, b) and narrows as coefficients are reduced further (c, d), though anti-diagonal pruning retains a consistent edge even at the lowest complexity point. This confirms that, for a fixed coefficient count, selectively pruning the least significant basis functions preserves linearization performance better than uniformly truncating the model order, supporting the pruning strategy adopted in SCALE-X.

\begin{figure*}[!t]
\centering
\includegraphics[width=7.2in]{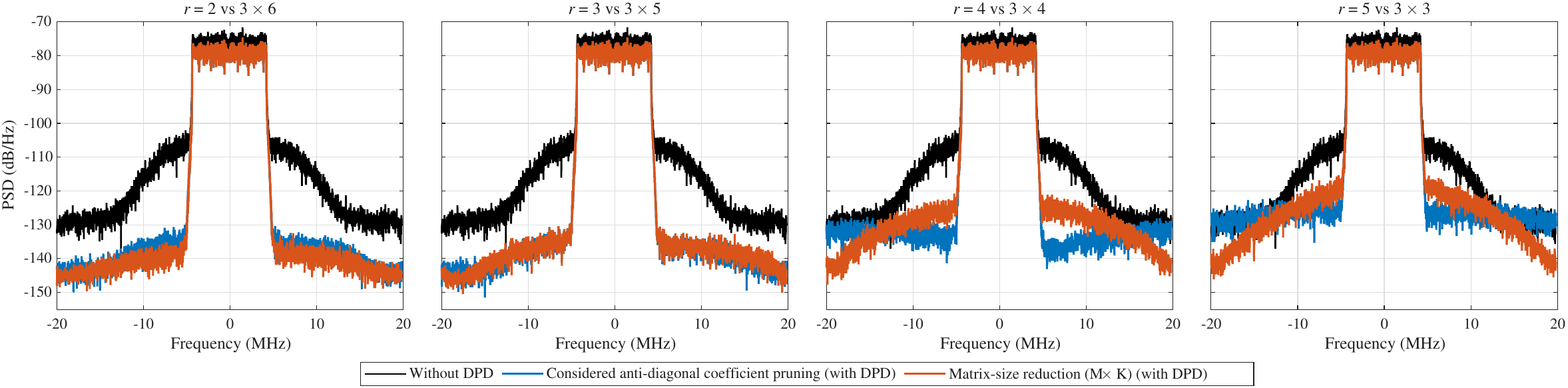}
\caption{Power spectral density of PA output for matched-complexity DPD configurations: anti-diagonal  constrained LS (blue) vs. $M\times K$ reduction (orange), without DPD PA output (black) as reference for various $r$, $M$ and $K$ values.
}
\label{fig:prev_vs_perf}
\end{figure*}

Table~\ref{tab:fpga_comparison} reports the FPGA resource utilization and power consumption of the proposed method across pruning stages $r=0$ to $r=5$. The equivalent number of slices (ENS), defined in the table notes, condenses resource usage into a single metric. Its $\max$ term reflects that look-up tables (LUTs) and flip-flops (FFs) share the same slice, with weights corresponding to the soft-logic cost of replicating a 25$\times$18 DSP multiply--accumulate and a 36\,kb block RAM, respectively. A 7-series slice (4 LUT6, 8 FF) is adopted as the common unit for all designs, since native slice counts on UltraScale+ parts differ (8 LUT6, 16 FF). As the compared devices span 28\,nm (Zynq-7000 7Z010) and 16\,nm FinFET (ZU48DR, VU9P) process nodes with differing DSP primitives, ENS is read as an architecture-level comparison of logic quantity. Since sample rates differ across designs, ENS is normalized by throughput (TP), giving ENS/TP in slices per Msps.

As $r$ increases, progressive zeroing of anti-diagonal entries in $\mathbf{C}^{(r)}$ directly reduces active multiplications: DSP usage falls from 50 to 25 (50\% reduction), with LUTs and FFs following the same trend, from 1423 to 1168 and 718 to 471, respectively. This drives ENS down from 7856 at $r=0$ to 4042 at $r=5$ (48.6\% reduction), almost entirely through DSP elimination on the pruned anti-diagonals. Since throughput remains constant at 375\,Msps across all stages, ENS/TP falls by the same proportion, from 21.0 to 10.8. Power decreases correspondingly from 218 to 177\,mW, confirming tangible energy savings alongside resource reduction. BRAM utilization remains at zero throughout, as coefficient storage is not the limiting factor at this model size.

\begin{table}[t]
\centering
\caption{Performance Comparison of Existing FPGA-based DPDs.}
\label{tab:fpga_comparison}
\resizebox{\linewidth}{!}{%
\begin{tabular}{|l|c|c|c|c|c|c|}
\hline
\textbf{Design} & \textbf{$r^*$} & \textbf{Bitwidth} &
\textbf{LUT/FF/DSP/BRAM} & \textbf{Power} &
\textbf{TP} & \textbf{ENS/TP} \\
\hline\hline

\multirow{6}{*}{Proposed}
  & 0 & \multirow{6}{*}{12-bit}
  & 1423/718/50/0 & 218 & 375 & 21.0 \\
\cline{2-2}\cline{4-7}
  & 1 & & 1415/694/48/0 & 211 & 375 & 20.1 \\
\cline{2-2}\cline{4-7}
  & 2 & & 1372/653/43/0 & 209 & 375 & 18.1 \\
\cline{2-2}\cline{4-7}
  & 3 & & 1285/577/37/0 & 187 & 375 & 15.7 \\
\cline{2-2}\cline{4-7}
  & 4 & & 1220/527/31/0 & 185 & 375 & 13.2 \\
\cline{2-2}\cline{4-7}
  & 5 & & 1168/471/25/0 & 177 & 375 & 10.8 \\
\hline\hline
\cite{huang2019parallel}
  & --- & 16-bit
  & 1826/2815/106/0
  & 185 & 400 & 40.9 \\
\hline

\cite{jiang2026lowcomplexity}
  & --- & ---
  & 20495/24710/504/108
  & --- & 625 & 154.0 \\
\hline

\cite{versluis2025sparsedpd}
  & --- & 14-bit
  & 2298/1724/66/13
  & 241 & 170 & 72.6 \\
\hline

\multirow{2}{*}{\cite{li2022fpga}$^{*}$}
  & \multirow{2}{*}{---} & \multirow{2}{*}{18-bit}
  & 11254/20318/622/5$^{\dag}$ & \multirow{2}{*}{5000} & 431.4 & 224.5 \\
\cline{4-4}\cline{6-7}
  &  & & 6469/92662/796/0$^{\ddag}$ & & 709.7 & 184.6 \\
\hline

\end{tabular}%
}

\vspace{1mm}
\parbox{\linewidth}{\footnotesize
\textit{Notes:}
Device platforms for \cite{huang2019parallel},
\cite{jiang2026lowcomplexity}, \cite{versluis2025sparsedpd}, and
SCALE-X are the Xilinx MPSoC Evaluation Board, Xilinx
xcvu9p-flga2577-2-i, Xilinx Zynq-7Z010, and Xilinx Zynq UltraScale+
RFSoC ZU48DR, respectively. Power is in mW; throughput (TP) is in
Msps. LUT/FF/DSP/BRAM denote FPGA resource counts. ENS is defined as
$\text{ENS} = \max(\text{LUT}/4,\text{FF}/8) + 150\cdot\text{DSP} +
144\cdot\text{BRAM}$, and ENS/TP is a normalized metric. $^{*}$:  reports a combined system (envelope generator, leakage canceller, and DPD); DPD-only resource usage is not separately reported,  $^{\dag}$: HLS $^{\ddag}$: Custom RTL. 
}
\end{table}

\subsection{Performance Comparison}
Table~\ref{tab:fpga_comparison} benchmarks the proposed design against four state-of-the-art FPGA DPD implementations~\cite{huang2019parallel,jiang2026lowcomplexity,versluis2025sparsedpd,li2022fpga}. Implemented on the Zynq UltraScale+ RFSoC ZU48DR at 12-bit (Q1.11) precision, the pruned design at $r^*=5$ uses 25 DSPs and no BRAMs, draws 177\,mW at 375\,Msps, and achieves the lowest ENS/TP (10.8).

The parallel architecture in~\cite{huang2019parallel} runs at 400\,Msps but needs 106 DSPs (ENS/TP of 40.9). The sparsity-driven design in~\cite{versluis2025sparsedpd} has a moderate footprint, but its 170\,Msps throughput yields 72.6. The design in~\cite{jiang2026lowcomplexity} has the largest DPD-only footprint (504 DSPs, 108 BRAMs), giving 154.0. The work in~\cite{li2022fpga} reports a combined envelope-generation, leakage-cancellation, and DPD system without separate DPD-only usage; it requires 622--796 DSPs depending on the design flow (HLS vs. custom RTL), with ENS/TP of 224.5 and 184.6, respectively.

Relative to~\cite{huang2019parallel}, the proposed design uses about $4\times$ fewer DSPs at comparable throughput, with $3.8\times$ lower ENS/TP and lower power (177 vs. 185\,mW). It reduces ENS/TP by $6.7\times$ and $14.3\times$ relative to~\cite{versluis2025sparsedpd} and~\cite{jiang2026lowcomplexity}, respectively, although~\cite{jiang2026lowcomplexity} runs at a higher throughput. Against~\cite{li2022fpga}, the comparison should be read with the scope difference in mind, since that work includes a full multi-stage system. Overall, the proposed method offers the most favorable balance of resource usage, power, and throughput for resource-constrained DPD deployment.

\section{Conclusion}\label{SecV}
SCALE-X presented a systematic, complexity-aware, low-precision, model-based DPD framework that combined coefficient pruning with optimized FxP modeling to capture quantization effects and guide efficient design. By jointly optimizing model order, memory depth, and coefficient bitwidth, it significantly reduced computational complexity while preserving linearization performance. FPGA results confirmed reduced hardware resource usage with improved hardware efficiency. 


\bibliographystyle{IEEEtran}
\bibliography{ref}

\end{document}